\documentclass{article}
\usepackage{spconf,amsmath,graphicx}
\usepackage{bm}
\usepackage{booktabs}
\usepackage{multirow}
\usepackage{array}
\usepackage{makecell}
\usepackage{hyperref}
\usepackage[
backend=biber,
style=numeric,
sortcites,
sorting=none,
giveninits=true,
maxbibnames=99,
doi=false,isbn=false,url=false,eprint=false
]{biblatex}
\hypersetup{
     colorlinks=true,
     citecolor=blue,
}
\newcommand\blfootnote[1]{%
  \begingroup
  \renewcommand\thefootnote{}\footnote{#1}%
  \addtocounter{footnote}{-1}%
  \endgroup
}

\title{A Preliminary Investigation on Flexible Singing Voice Synthesis Through Decomposed Framework with Inferrable Features}
\name{Lester Phillip Violeta*$^{1}$, Taketo Akama$^{2}$}
\address{$^1$Nagoya University, Japan, $^{2}$Sony Computer Science Laboratories, Inc., Tokyo, Japan  
}

\begin{document}
%
\maketitle
\vspace{7.4em}
\begin{abstract}
We investigate the feasibility of a singing voice synthesis (SVS) system by using a decomposed framework to improve flexibility in generating singing voices. Due to data-driven approaches, SVS performs a music score-to-waveform mapping; however, the direct mapping limits control, such as being able to only synthesize in the language or the singers present in the labeled singing datasets. As collecting large singing datasets labeled with music scores is an expensive task, we investigate an alternative approach by decomposing the SVS system and inferring different singing voice features. We decompose the SVS system into three-stage modules of linguistic, pitch contour, and synthesis, in which singing voice features such as linguistic content, F0, voiced/unvoiced, singer embeddings, and loudness are directly inferred from audio. Through this decomposed framework, we show that we can alleviate the labeled dataset requirements, adapt to different languages or singers, and inpaint the lyrical content of singing voices. Our investigations show that the framework has the potential to reach state-of-the-art in SVS, even though the model has additional functionality and improved flexibility. The comprehensive analysis of our investigated framework's current capabilities sheds light on the ways the research community can achieve a flexible and multifunctional SVS system.
\end{abstract}
\section{Introduction}
\blfootnote{*Work was conducted at Sony Computer Science Laboratories, Inc., Tokyo.}
\label{sec:introduction}
Singing voice synthesis (SVS) \cite{oura2010recent, hono2018recent} is the task of generating natural and expressive singing voices given musical score inputs, which contain the text\footnote{In this work, text primarily refers to the lyrics of the song.}, MIDI notes, and the phoneme/MIDI and audio alignments. Synthesizing singing voices is commonly considered the intersection between the music and speech information processing fields, taking into consideration ideas from both areas to develop high-quality systems. Compared to speech, singing voices are more difficult to model due to the presence of high-frequency components and the multiple possible variations in singing the same text. However, owing to the rise of neural networks, data-driven methods have enabled the synthesis of high-quality singing voices. 

One main difficulty in SVS is the collection of labeled training data. A typical SVS dataset typically contains the singing waveform and the corresponding music score. As current frameworks are data-driven, the training datasets need to contain audio samples of the target singer, and singing in the language that the user wants to synthesize in. However, collecting and labeling singing datasets is a very tedious task, and can quickly become expensive to do with new singers and languages. Without such a dataset, SVS functionalities become limited to synthesizing in a language or singer in a small dataset. In this paper, we investigate SingFlex, which resolves the limited capability issues caused by stringent dataset requirements. 

\begin{figure}[t!]
  \centering
  \includegraphics[width=5.2cm]{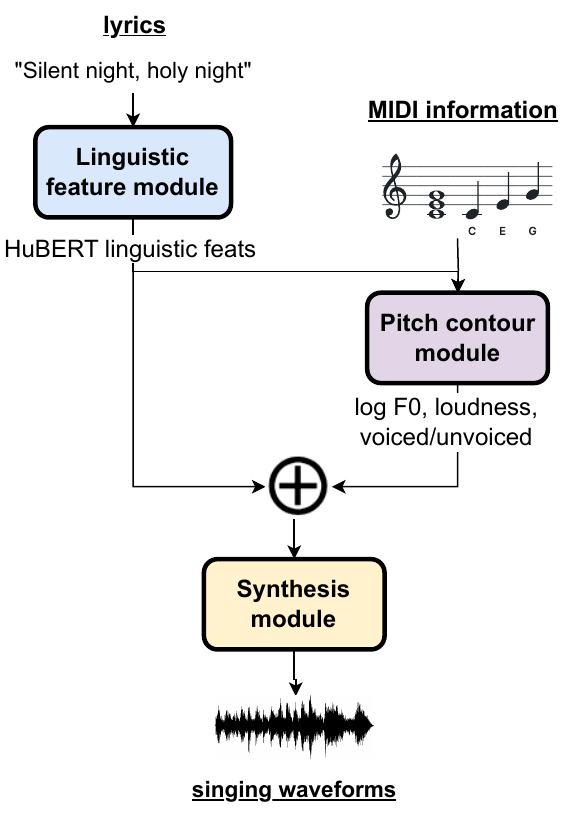} 
  \caption{An overview of the investigated method SingFlex. The text are transformed into HLFs, which are used to predict the pitch and other acoustic features. From these features, the singing waveforms are synthesized.}
  \label{fig:overview}
\end{figure}

Through a decomposed framework, as illustrated in Fig. \ref{fig:overview}, we show that we can enable several flexible SVS functionalities, such as the following:

\noindent{\textbf{Alleviate the dataset label requirements.}}
This is achieved through the use of inferrable self-labeled features to train the pitch and synthesis networks. Aside from inferring F0, voiced/unvoiced (VUV), loudness, and singer embeddings from the singing voices, we also infer HuBERT linguistic features (HLFs) \cite{hsu2021hubert, hsoft-vc-2022}, a continuous time-aligned representation of discrete text. As HLFs have been made specifically for VC and cross-lingual tasks \cite{hsoft-vc-2022}, these have been seen as a strong choice as linguistic features for singing synthesis \cite{huang2023svcc}. Moreover, we use MIDI predictors to infer the MIDI information from the F0 and audio. Since we can also use unlabeled singing voice data to train the pitch and synthesis modules, these two modules can now handle various singing voice domains such as genres, techniques, singers, and languages, which was not possible before.

This approach changes conventional SVS approaches, which directly used linguistic features estimated from the text and MIDI labels from the music score labels to train the network. Thus, with the decomposed framework, all we need is a text labeled speech data (public datasets are widely available thanks to the extensive research in text-to-speech), which reduces the label requirements, drastically opening the possibility of synthesizing various singing voices to adopt to customized experiences from a user perspective.

\noindent{\textbf{Adapt to different languages.}}
This is done by swapping out the linguistic module to infer HLFs from text. With a decomposed framework and the HLFs being disentangled from melody-related information, we can train a linguistic module using a speech dataset to generate HLFs, resulting in the ability to generate singing voices in any language that the linguistic module is trained on. As speech datasets with text labels and in different languages are more available than singing datasets, this allows us to use a large-scale dataset to train the linguistic module.

\noindent{\textbf{Adapt to different singers.}}
This is achieved by conditioning the pitch and synthesis modules with the inferrable target singer’s embedding. Usually, to synthesize with a new singer, we need to label at least MIDI and text for the singer. However, through the self-labeled training method including the MIDI and text related feature labels, we can use a large multi-singer singing dataset to train a multi-singer synthesis network without the need of labeling all of these with music scores.

\noindent{\textbf{Inpaint the lyrical content of singing voices.}}
This is accomplished by manipulating the HLFs used by the synthesis module, such as concatenating the HLFs inferred from audio and HLFs estimated by the linguistic module. Thus, the lyrical content of a singing waveform can be inpainted by masking the inferred HLFs, and replacing the masked segments with HLFs inferred by the linguistic module with the new lyrics. The audio with the inpainted lyrical content can be resynthesized by using the resulting concatenated HLFs by only needing to provide the text of the segment to be inpainted, without explicitly estimating the text of the context audio, which avoids solving unnecessarily complex tasks. Moreover, since HLFs can extract linguistic features in different languages, this can enable multi-lingual SVS without needing an audio sample of a singer singing in two different languages.

We summarize the contributions of this paper in the following points:
\begin{itemize}
  \item We investigate SingFlex, a decomposed framework that uses inferrable features, improving the flexibility in SVS such as alleviating the labeled dataset requirements, adapting to different languages or singers, and inpainting the lyrical content of singing voices. This leads to the development of a model that can support various singing voice domains such as genres, techniques, singers, and languages, providing users with a personalized SVS system that allows for editing.
  \item The modules of our framework include two key proposals: the first is a simple yet effective approach to improve the MIDI inference from singing voice, and the second is an efficient model to quickly infer linguistic features.
  \item Our results indicate that our SingFlex has the potential to reach state-of-the-art in SVS, while incorporating enhanced functionality and improved flexibility. Our analysis of its current capabilities reveals pathways for the research community to establish a genuinely flexible and multifunctional system.
\end{itemize}
We recommend that readers refer to the samples available on our demo page\footnote{\url{singflex.github.io}}.

\begin{figure*}[t!]
  \centering
  \includegraphics[width=16.5cm]{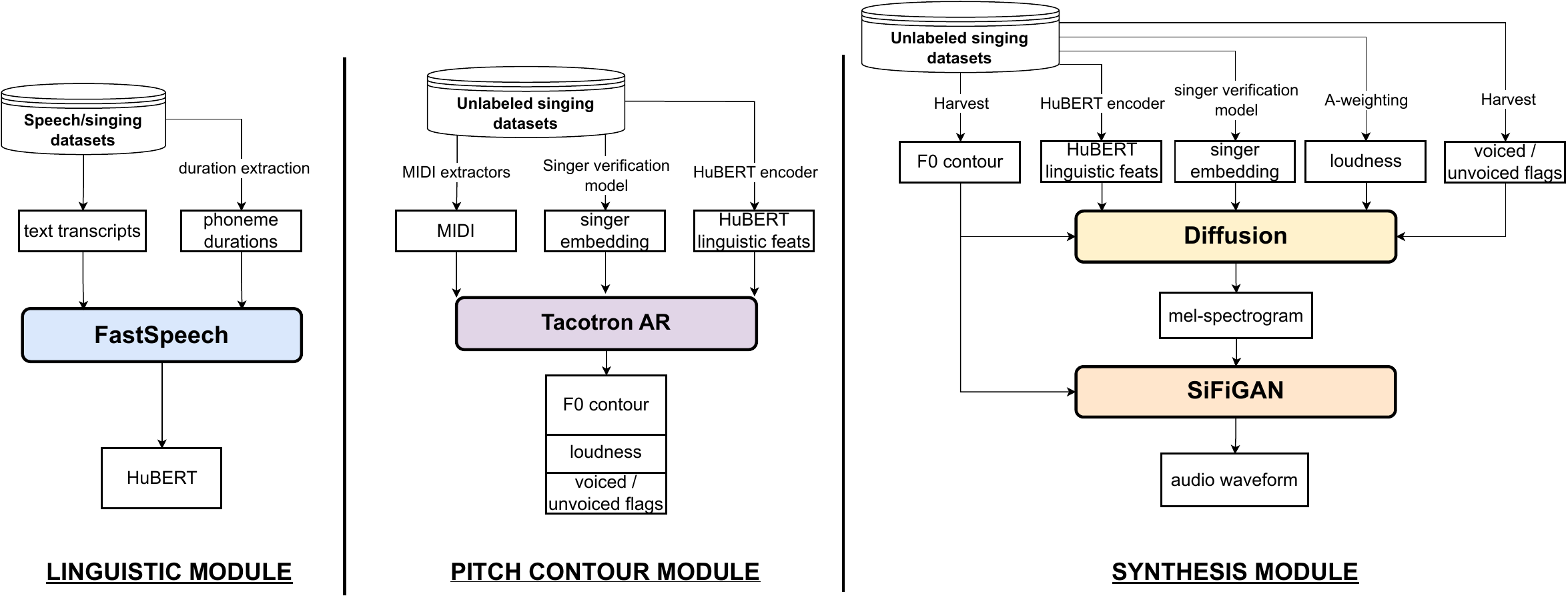} 
  \caption{Detailed visualizations of each module and the datasets used to train each module.}
  \label{fig:archi}
\end{figure*}

\section{Related literature}
\label{sec:related}
\subsection{Singing synthesis from music score inputs}
\label{sec:related_svs}
A common application of singing synthesis is to synthesize a singing voice given the music score information through an acoustic model. Sinsy \cite{hono2021sinsy} transforms the music score into HTS full-context labels, a continuous representation of the music score. As human singers typically do not strictly follow the music score, Sinsy proposed a time-lag (which models the difference of the alignment between the music score and the singing voice) and a duration model (which models the lengths of each phoneme with consideration to the notes), to modify the continuous representations before being processed by an acoustic model. Other methods have also been proposed to improve the performance of acoustic modeling. For example, \cite{blaauw2017neural, blaauw2020sequence} proposed multi-stream modeling to improve the correlation between the different acoustic features. Here, the acoustic features are decomposed into the log F0, mel-spectrum, and aperiodicity. Each feature is modeled by a different network; however, the networks are cascaded together such that the predicted features are also used as inputs to predict another feature, making the features correlated to each other. NNSVS \cite{yamamoto2023nnsvs}, an open-sourced toolkit with implementations of the two aforementioned modularized SVS methods, has investigated these two frameworks thoroughly, and showed high synthesis naturalness particularly by synthesizing out of range pitches without degradations.

However, these aforementioned data-driven approaches have several limitations in use cases which are restricted by the dataset labels. For example, synthesizing a singing voice in another singer's voice not in the training data is not possible. Another limitation is generating in another language outside the training data. While both cases can be resolved by collecting a new dataset and training multiple models, the monetary and time costs can become very expensive quickly. Previous works such as \cite{wu2022_zerosvs} propose a unified model for synthesizing singers in zero-shot settings, and our work extends on this by including the ability to synthesize in different languages and with an increased inpainting functionality. Other works such as \cite{zhou2023bisinger} propose a method to synthesize bilingual singing voices through learning a shared phoneme representation of the two target languages. However, scaling this method to multi-lingual settings is difficult, as different languages have different phonemes.

\subsection{Singing synthesis from audio inputs}
\label{sec:related_svc}
Singing synthesis can also be done given audio inputs, where the end goal is usually to convert the singer's identity through singing voice conversion (SVC) \cite{huang2023svcc}. The mainstream framework in SVC has been a recognition-synthesis framework \cite{liu2021diffsvc, huang2022comparative}, where the linguistic representations of the singing voice are first encoded as a linguistic feature and are then used to synthesize the singing voice in the target singer's voice by conditioning the network with the target singer embedding. Generative models such as variational autoencoders (VAEs) \cite{unsupervised-svc, ning2023vits, zhou2023vits}, generative adversarial networks (GANs) \cite{ucd-svc, liu2021fastsvc, zhou2022hifi}, or diffusion models \cite{liu2021diffsvc, yamamoto2023svcc} have been used to generate the acoustic features or audio waveforms of the singer in the target singer given linguistic and pitch conditioning features. By using a strong generative model and a large-scale unlabeled dataset, a system should be able to generate waveforms of a specified target singer. 

Research works such as \cite{hsoft-vc-2022} have also shown that by fine-tuning the HuBERT model \cite{hsu2021hubert} to produce "soft" features, to disentangle the learned audio representations from HLFs. This has given path to the recognition-synthesis voice conversion (VC) frameworks \cite{huang2022comparative}, where VC systems use SSL encoders like HuBERT to extract dense linguistic information through HLFs through a recognition step and subsequently change the speaker information during the synthesis step. Moreover, in the recent SVCC 2023, top performing systems used HLFs (or one of its variants) as the recognizer, showing its potential as a linguistic encoder.

\section{Proposed Method}
Inspired by Sinsy-based methods which use HTS full-context labels described in Section \ref{sec:related_svs}, we instead use HLFs as the continuous time-aligned input representations. Compared to HTS full-context labels which are extracted from the music score, HLFs are extracted from the audio itself. Moreover, through the use of MIDI extractors, we can simply extract the MIDI note information from the audio itself. Through this framework, the functionalities for SVS can be improved without needing a large-scale labeled dataset. Our proposed SVS model is a three-stage framework explained in the following subsections. A detailed view of each module's architecture can be seen in Fig. \ref{fig:archi}.

\subsection{Linguistic feature module: HLF generation from time-aligned text inputs}
\label{sec:linguistic_module}
To generate new linguistic features from the text, we adopt the FastSpeech \cite{ren2019fastspeech} architecture, a commonly used framework in text-to-speech. FastSpeech's architecture is perfect for our task due to its non-autoregressive method of aligning the text to the audio. The model is trained by taking in the text phonemes and the duration vector of the text. The text phonemes are acquired from a grapheme-to-phoneme converter, and the duration vector represents how many frames are allotted for each phoneme, which is either derived from forced-alignment tools or the attention map from the attention map of a pretrained autoregressive model. In FastSpeech, a phoneme duration predictor is jointly trained with the mel-spectrogram model to predict the duration vector, such that the user only needs to input the text during inference.

To train the model, a discrete phoneme vector $X=[x_0,...,x_N]$ and a scalar duration vector of $D=[d_0,...,d_N]$, both of length $N$, where the sum of the values of ${D}$ is equal to the length $T$ of the target feature ${Y}=[\bm{y}_0,...,\bm{y}_T]$, are used as inputs. FastSpeech first processes the phoneme vector through a Transformer encoder ($\text{Enc}$) \cite{transformer} and are transformed to a feature ${H}=[\bm{h}_0,...,\bm{h}_N]$ in the attention dimension with feature lengths of size $N$ equal to the input. The encoder outputs ${H}$ are then expanded using a length regulator ($\text{LR}$), which repeats each value of ${H}$ using the duration vector ${D}$, resulting in ${I}=[\bm{i}_0,...,\bm{i}_T]$, a time-aligned feature to the target features ${Y}$, essentially matching each phoneme to its corresponding number of audio frames. The time-aligned features are then processed by another Transformer encoder ($\text{Dec}$) to produce the output target features ${Y}$. 

To adopt the FastSpeech architecture to our task, we simply remove the phoneme duration predictor and just use the ground truth durations, as different from speech, the phoneme durations in singing voices are constrained by the music score, or can be manually modified by the user. Moreover, we predict the HLFs \cite{hsu2021hubert, hsoft-vc-2022} instead of log mel-spectrograms, which will be used as inputs in the pitch contour and synthesis modules. In the language adaptation task, we simply swap out the linguistic module with another one trained on a different language.

\subsection{Pitch contour module: F0 generation from MIDI inputs}
\label{sec:pitch_module}
The pitch contour module generates the pitch contours given the predicted HLFs and the MIDI information. As predicting an F0 sequence is a difficult task, we guide the F0 prediction through residual log F0 modeling \cite{hono2021sinsy, lu2020xiaoicesing}, which has been an effective method in SVS with autoregressive models \cite{blaauw2017neural, yi2019singing}. Residual log F0 modeling uses the input MIDI information as a guide and predicts the bias between the MIDI and the ground truth pitch contours effectively simplifying the F0 generation task. As mentioned in Section \ref{sec:introduction}, since it is difficult to have the groundtruth MIDI labels, we need to find to acquire these MIDI information from just the audio itself. To resolve this issue, we extract the pitch contour from the audio using Harvest \cite{morise2017harvest} and flatten the pitch contour to acquire the time-aligned MIDI information. This is done by using two MIDI extractors, one based on phoneme information \cite{2023yongphonememidi}, represented as ${P}=\{\bm{p}_0,...,\bm{p}_i\}$ and another trained on polyphonic data \cite{polyphonic2022midi}, represented as ${Q}=\{\bm{q}_0,...,\bm{q}_i\}$. To create the flattened MIDI representation ${M}=\{\bm{m}_0,...,\bm{m}_i\}$ The results from both are combined such that each frame from the MIDI extractors are compared to the Harvest pitch contour, and the frame closer in value to the pitch is used as the MIDI information, as seen in the formulation below.

\begin{equation}
    {m}_i = \begin{cases} 
    q_i, & \text{if } |h_i - p_i| \geq |h_i - q_i| \\
    p_i, & \text{otherwise}
\end{cases}
\end{equation}

We then use an autoregressive Tacotron-based model \cite{shen2018taco2} as the pitch contour module to predict the residual log F0. To synthesize in a target singer, we use a pretrained singer verification model trained on singing data \cite{torres2023sslembeddings} and add it to the outputs of the encoder to condition the module. To summarize, the pitch contour module takes in the flattened MIDI and linguistic features as inputs to produce the F0 pitch contour, VUV features, and loudness features.

\subsection{Singing synthesis module: Singing waveform generation from linguistic, pitch, and timbre features}
\label{sec:synthesis_module}
After the linguistic module predicts the HLFs and the pitch contour module predicts the F0 information, we use these as conditioning features to produce the singing waveforms through a generative model, similar to the recognition-synthesis module discussed in Section \ref{sec:related_svc}. We use a Diffusion model \cite{ho2020denoising, liu2021diffsinger} as its backbone. To train the model, noise is iteratively added to the target log mel-spectrogram for $N$ timesteps, and then the loudness, log F0, VUV, and HLFs are used as conditioning features to predict the noise from the mel-spectrogram at timestep $N+1$. During inference, Gaussian noise is used as input and the mel-spectrogram is predicted after $N$ denoising iterations. Finally, to synthesize the audio waveforms from the predicted mel-spectrograms, a separately trained vocoder is used as an inverter.

Specifically, we adopt the non-causal WaveNet \cite{liu2021diffsinger} for the Diffusion model with 20 layers of one-dimensional residual connections. We use the variant called HuBERT soft \cite{hsoft-vc-2022} to extract HLFs, which further disentangles the features from speaker-related features and performs well on multi-lingual data. For the vocoder, we adopt SiFiGAN \cite{yoneyama23sifigan} which additionally uses F0 information to generate audio waveforms from mel-spectrograms. Similar to the pitch contour module, we use a pretrained singer verification model trained on singing data \cite{torres2023sslembeddings} to synthesize in a target singer. We add the singer embeddings along with the time embeddings at every residual block output of the Diffusion model.

\subsection{Inpainting procedure}
\label{sec:inpainting_procedure}
As HLFs are essentially a continuous representation of discrete text, we demonstrate how this can be simply manipulated and enable inpainting tasks such as lyric editing and multilingual singing synthesis. In the case of mixing audio and text/MIDI inputs, we simply concatenate the HLFs, log F0, VUV, and loudness extracted from the audio and the HLFs, log F0, VUV, and loudness predicted by the linguistic and pitch contour modules. Then, the resulting concatenated HLFs can be used as inputs to the synthesis module to generate the inpainted singing waveforms with more control in the text content.

\section{Experimental Setup}
\begin{table}[t]
\centering
    \small
    \caption{Details of the datasets used to train the pitch and synthesis modules.}
    \label{tab:datasets}
\begin{tabular}{ccccccc}
\toprule
\textbf{Dataset} & \textbf{Language} & \textbf{Hours} & \textbf{Singers} \\
\midrule

Namine Ritsu \cite{2020ritsu_enunuv2} & Japanese & 4.35 & 1 \\
Tohoku Kiritan \cite{ogawa2021tohoku} & Japanese & 0.95 & 1 \\
PJS \cite{koguchi2020pjs} & Japanese & 0.45 & 1 \\
Itako \cite{itako} & Japanese & 0.43 & 1 \\
Natsume Yuri \cite{natsume} & Japanese & 0.43 & 1 \\
JSUT-Song \cite{sonobe2017jsut} & Japanese & 0.37 & 1 \\
M4Singer \cite{m4singer} & Mandarin & 29.7 & 20  \\
OpenCPop \cite{opencpop} & Mandarin & 5.23 & 1 \\
CSD \cite{choi2020csd} & Korean & 2.23 & 1 \\
\bottomrule
\end{tabular}
\end{table}

\subsection{Training and inference details}
We synthesize singing voices at 44.1 kHz sampling rate. We extract all features with a hop size of 220. We train both the linguistic module, pitch contour module, and the Diffusion model in the synthesis module for up to 300 epochs. For the SiFiGAN vocoder, we train it for 250 epochs. During inference, we randomly take a separate singing sample of the target singer to extract the singing embedding. The Diffusion model denoises the Gaussian noise inputs for 100 steps. In cases where the target singer is also converted, we also shift the key of the input MIDI using a mean variance transformation.

\subsection{Baseline comparison}
We use NNSVS \cite{yamamoto2023nnsvs} as our baseline, which is an open-sourced toolkit, and has implementations of state-of-the-art SVS models. We use the recently developed recipe\footnote{\url{https://github.com/nnsvs/nnsvs/tree/master/recipes/namine_ritsu_utagoe_db/dev-48k-melf0}} which predicts log F0, and VUV information from the input music score using an autoregressive model, and the mel-spectrograms from the input music score, along with the predicted, using a Diffusion model \cite{liu2021diffsinger}. We use the same SiFiGAN vocoder described in Section \ref{sec:synthesis_module} to invert the mel-spectrograms and F0 information into audio waveforms. We also change the recipe configuration to synthesize at a 44.1 kHz sampling rate and a hop size of 220 from the original 48 kHz and hop size of 240.

\begin{table*}[h!]
  \centering
    \caption{Summary of evaluation results. Note that only setups with ground truth references were evaluated for MCD, F0 RMSE, and F0 CORR objective metrics. The MOS results are presented with a 90\% confidence interval.}
    \label{tab:main_results}
    \small
\begin{tabular}{lllccccccc}
\toprule
\textbf{Sys.} & \textbf{Language} & \textbf{Description} & \textbf{MCD~($\downarrow$)} & \textbf{F0 RMSE~($\downarrow$)} & \textbf{F0 CORR~($\uparrow$)} & \textbf{CER/WER~($\downarrow$)} & \textbf{MOS~($\uparrow$)} \\
\midrule
1 & \multirow{5}{*}{\begin{tabular}[c]{@{}l@{}}In-domain\\ (Japanese)\end{tabular}} & NNSVS \cite{yamamoto2023nnsvs} & 8.42 & 41.63 & 0.83 & 20.14 & 3.90 $\pm$ 0.38 \\
2 & & \textbf{SingFlex (Proposed)} & 8.85 & 51.50 & 0.76 & 21.88 & 3.62 $\pm$ 0.44 \\
3 & & \hspace{0.4 cm} w/o linguistic module & 7.68 & 38.89 & 0.84 & 25.82 & 3.67 $\pm$ 0.48 \\ 
4 & & \hspace{0.4 cm} w/o large-scale pretraining & 8.97 & 54.09 & 0.75 & 27.73 & 2.13 $\pm$ 0.34 \\ 
5 & & \hspace{0.4 cm} w/ singer conversion & - & - & - & 21.31 & 2.13 $\pm$ 0.34 \\ 
\cmidrule(lr){1-3}
6 & \multirow{3}{*}{\begin{tabular}[c]{@{}l@{}}Out-of-domain\\ (English)\end{tabular}} & \hspace{0.4 cm} w/ speech linguistic module & - & - & - & 30.50 & 2.00 $\pm$ 0.36 \\ 
7 & & \hspace{0.4 cm} w/o linguistic module & - & - & - &  6.78 & 3.05 $\pm$ 0.48 \\
8 & & \hspace{0.4 cm} w/ bilingual inpainting & - & - & - & - & 2.18 $\pm$ 0.35 \\
\midrule
9 & Japanese & Ground truth & - & - & - & 18.75 &  4.53 $\pm$ 0.26 \\ 
10 & English & Ground truth &  - & - & - & 6.78 & 4.65 $\pm$ 0.26 \\ 
\bottomrule
\end{tabular}
\end{table*}

\subsection{Datasets}
To evaluate SingFlex in an SVS task, we used the Namine Ritsu dataset \cite{2020ritsu_enunuv2}, which is sung by a single singer in Japanese of 110 songs totaling to around 4.35 hours, and used the same train/dev/test split of 100/5/5 provided in NNSVS. To evaluate the performance of the model in an unseen singing language like English, we used the KENN04 song of the NUS-48E dataset \cite{nus48e}. The NUS-48E dataset contains the phoneme and audio alignments, but not the MIDI information, thus we use the MIDI extraction method described in Section \ref{sec:pitch_module} during evaluation in Section \ref{sec:results_languages} and Section \ref{sec:results_inpainting}. During inpainting in Section \ref{sec:results_inpainting}, we change half of the segments into its Japanese lyrics by considering the MIDI to enable bilingual singing synthesis.

For the linguistic module, we took advantage of the relaxed training data capabilities of SingFlex, and pretrained on larger datasets before fine-tuning on the Namine Ritsu dataset. We pretrained the text encoder on the JSUT dataset, which contains 10 hours of Japanese speech data. The durations of each phoneme were extracted using the attention map from a pretrained autoregressive model as described in FastSpeech \cite{ren2019fastspeech}. For the language adaptation task, we trained another linguistic module, using the 100-hr and 360-hr subsets of LibriTTS \cite{zen2019libritts}, an English speech dataset. The durations of each phoneme were extracted using the Montreal Forced Aligner \cite{mcauliffe17_mfa}, a commonly used forced aligner tool in speech processing. Note that we did not fine-tune the linguistic encoder on the NUS-48E singing dataset to explore its performance being trained only with speech data.

For the pitch and synthesis modules, we collected several multi-lingual singing datasets for large-scale pretraining, which are described in detail in Table \ref{tab:datasets}. Note that although most of these datasets have MIDI information included, we did not use them and used the MIDI information extraction technique described in Section \ref{sec:pitch_module} from the audio to train the model and that we did not include any English singing data in this dataset. We used this pretrained model for synthesizing in an unseen language like English, but fine-tuned this model on the Namine Ritsu dataset for the Japanese task.

\subsection{Evaluation methods}
For subjective tests, we conducted the mean opinion score (MOS) test, primarily considered as the gold standard for evaluating synthesized singing voices. We recruited 15 evaluators who speak both Japanese and English and ask them to rate each sample from 1 to 5, with 1 being the lowest and 5 being the highest. We took 4 random samples from each system to be evaluated.

For objective tests, we used the character/word error rate (CER/WER) to verify the intelligibility, and which was also the objective metric most correlated to human percepted naturalness, as seen in SVCC 2023 \cite{huang2023svcc}. To calculate the CER in Japanese, we used a Conformer-based architecture \footnote{\url{https://huggingface.co/reazon-research/reazonspeech-nemo-v2}}. To calculate the WER in English, we used Whisper \cite{radford2023whisper}. To measure how much the singing style matches with the ground truth, we also used mel-cepstral distortion (MCD) and F0-related objective metrics such as the F0 root mean square error (F0 RMSE) and F0 correlation (F0 CORR).

\section{Results and Discussion}
\subsection{Comparison with SVS baseline}
\label{sec:results_svs}
We compare our SingFlex system with NNSVS. As seen in the results, our proposed system (\textbf{Sys. 2}) is comparable to NNSVS (\textbf{Sys. 1}) in terms of intelligibility through the CER objective metric (21.88\% vs. 20.14\%), but exhibits a minor reduction in naturalness through the MOS test (3.62 vs. 3.90). On the other hand, upon comparing the synthesized and ground truth samples, we see that NNSVS has better objective scores in MCD, F0 RMSE, and F0 CORR, showing that the singing style from the input MIDI may have not been truthfully copied in SingFlex.

However, when removing the linguistic module (by using the ground truth HLFs extracted from audio) in \textbf{Sys. 3}, the samples had better scores in MCD, F0 RMSE, and F0 CORR than NNSVS. Although it was expected to have better MCD scores due to the HLFs being inferred from audio itself, having better F0-related scores show that the pitch contour module can successfully synthesize the F0 from HLFs as conditioning features. This huge gap between predicted (\textbf{Sys. 2}) and ground truth HLFs (\textbf{Sys. 3}) in F0 scores also shows that although using HLFs has potential, more work needs to be done to improve the linguistic module to generate better HLFs.

\subsection{Alleviation of dataset label requirements}
Aside from the conventional SVS task, we further investigate several other functionalities of SingFlex which are not available in NNSVS. Comparing \textbf{Sys. 2} and \textbf{Sys. 4}, we see that a multifunctional model requires large-scale data, and thus \textbf{Sys. 2} successfully improved the performance from \textbf{Sys. 4} by integrating the large-scale pretraining of our proposed framework.

\subsection{Adaptation to different languages}
\label{sec:results_languages}
We investigate SingFlex's ability to synthesize singing voices in a different language by swapping the text encoder by synthesizing the KENN04 song in Namine Ritsu's singing voice. First, we observe that despite the pitch and synthesis modules not being trained on English, there are no performance WER degradations when removing the linguistic module (or when using the ground truth HLFs) in \textbf{Sys. 7} compared to the English ground truth samples (\textbf{Sys. 10}), showing the multi-lingual capability of HLFs.

Moreover, similar to the findings in Section \ref{sec:results_svs}, we see a performance degradation when using the predicted HLFs (\textbf{Sys. 6}), with almost a 24\% relative degradation in CER and 1.07 point MOS degradation compared to \textbf{Sys. 7}, which could mainly be attributed to the linguistic encoder being trained only on English speech. In particular, we observe that some phonemes become whispered. This may come from the fact that although HLFs are disentangled from melody information, there is still a gap in phoneme durations between singing and speech, showing that methods such as using singing data or duration augmentated speech is still needed to train the linguistic module and make it perform similar to state-of-the-art baselines like NNSVS. 

\subsection{Adaptation to different singers}
\label{sec:results_inpainting}
Moreover, we also convert to another singer from the PJS dataset \cite{koguchi2020pjs} in \textbf{Sys. 5}. We see that SingFlex synthesizing in a different singer is also comparable to the Namine Ritsu setup in terms of intelligibility (21.88\% vs. 21.31\%); however, we observe some errors in the VUV predictions, causing some degradations in the MOS naturalness score. This may be attributed to using the pretrained multisinger pitch contour and synthesis modules, showing that predicting the VUV features may still be a difficult task in a multisinger setting and that fine-tuning the pitch and synthesis modules further to a single singer is still needed, as this was not observed in the Namine Ritsu setup.

\subsection{Inpainting lyrical content}
Moreover, the manipulation of HLFs enables the bilingual inpainting task. Compared to the other setups, we see that \textbf{Sys. 8} has a lower naturalness score. Since we essentially just resynthesize the unmasked parts and know that from \textbf{Sys. 7} that the system can perform well in resynthesis, we observe that the degradation comes from the inpainted segments in the predicted HLFs.

\section{Conclusions}
We investigated SingFlex, an SVS system by decomposing the framework to improve controllability functionalities. Through this decomposed framework, we showed that we can alleviate the labeled dataset requirements, adapt to different languages or singers, and inpaint the lyrical content of singing voices, improving the functionality of SVS and creating customized experiences from a user perspective. We showed that SingFlex has the potential to reach state-of-the-art in SVS, and that the model has additional functionality and improved flexibility. We provided a comprehensive analysis of SingFlex's current capabilities, which provides insights on how the research community can achieve a flexible and multifunctional SVS system. 

\section{Future Directions}
Future work includes the improvement of the quality of the currently investigated system over the baseline and in the different investigated tasks. While the experiments showed the feasibility of the ideas, more work needs to be done to make the performance comparable with several other baseline SVS systems.

\section{Ethics Statement}
In today's age of neural networks, several frameworks can now generate natural-sounding samples of any singer given their singing data. Using systems such as SingFlex are helpful and provide entertainment, but is a double-edged sword, as they can also be used to manipulate text and create vulgar content, or generate new songs without properly compensating the singer the model emulates. Thus, researchers need to be cautious in using these models for their use cases.

\section{References}
\printbibliography

@inproceedings{hono2018recent,
  title={Recent development of the {DNN}-based singing voice synthesis system—sinsy},
  author={Hono, Yukiya and Murata, Shumma and Nakamura, Kazuhiro and Hashimoto, Kei and Oura, Keiichiro and Nankaku, Yoshihiko and Tokuda, Keiichi},
  booktitle={Proc. APSIPA},
  pages={1003--1009},
  year={2018},
}

@article{hono2021sinsy,
  title={Sinsy: A deep neural network-based singing voice synthesis system},
  author={Hono, Yukiya and Hashimoto, Kei and Oura, Keiichiro and Nankaku, Yoshihiko and Tokuda, Keiichi},
  journal={IEEE/ACM Trans. on Audio, Speech, and Lang. Process.},
  volume={29},
  pages={2803--2815},
  year={2021},
  publisher={IEEE}
}

@inproceedings{oura2010recent,
  title={{Recent development of the HMM-based singing voice synthesis system—Sinsy}},
  author={Oura, Keiichiro and Mase, Ayami and Yamada, Tomohiko and Muto, Satoru and Nankaku, Yoshihiko and Tokuda, Keiichi},
  booktitle={Seventh ISCA Workshop on Speech Synthesis},
  year={2010}
}

@article{blaauw2017neural,
  title={A neural parametric singing synthesizer modeling timbre and expression from natural songs},
  author={Blaauw, Merlijn and Bonada, Jordi},
  journal={Applied Sciences},
  volume={7},
  number={12},
  pages={1313},
  year={2017},
  publisher={MDPI}
}

@inproceedings{lu2020xiaoicesing,
  author={Peiling Lu and Jie Wu and Jian Luan and Xu Tan and Li Zhou},
  title={{XiaoiceSing: A High-Quality and Integrated Singing Voice Synthesis System}},
  year=2020,
  booktitle={Proc. Interspeech},
  pages={1306--1310},
}

@inproceedings{blaauw2020sequence,
  title={Sequence-to-sequence singing synthesis using the feed-forward transformer},
  author={Blaauw, Merlijn and Bonada, Jordi},
  booktitle={Proc. IEEE ICASSP},
  pages={7229--7233},
  year={2020},
}

@article{liu2021diffsinger,
  title={{DiffSinger: Singing voice synthesis via shallow diffusion mechanism}},
  author={Liu, Jinglin and Li, Chengxi and Ren, Yi and Chen, Feiyang and Liu, Peng and Zhao, Zhou},
  journal={AAAI},
  volume={36},
  number={10},
  pages={11020--11028},
  year={2022}
}

@inproceedings{yi2019singing,
  author={Yuan-Hao Yi and Yang Ai and Zhen-Hua Ling and Li-Rong Dai},
  title={{Singing Voice Synthesis Using Deep Autoregressive Neural Networks for Acoustic Modeling}},
  year=2019,
  booktitle={Proc. Interspeech},
  pages={2593--2597},
}

@INPROCEEDINGS{yamamoto2023nnsvs,
  author={Yamamoto, Ryuichi and Yoneyama, Reo and Toda, Tomoki},
  booktitle={Proc. IEEE ICASSP}, 
  title={{NNSVS: A Neural Network-Based Singing Voice Synthesis Toolkit}}, 
  year={2023},
  volume={},
  number={},
  doi={10.1109/ICASSP49357.2023.10096239}
}

@inproceedings{huang2023svcc,
  title={{The Singing Voice Conversion Challenge 2023}},
  author={Huang, Wen-Chin and Violeta, Lester Phillip and Liu, Songxiang and Shi, Jiatong and Yasuda, Yusuke and Toda, Tomoki},
  booktitle={Proc. ASRU},
  year={2023}
}

@inproceedings{yamamoto2023svcc,
  title={{A Comparative Study of Voice Conversion Models with Large-Scale Speech and Singing Data: The T13 Systems for the Singing Voice Conversion Challenge 2023}},
  author={Yamamoto, Ryuichi and Yoneyama, Reo and Violeta, Lester Phillip and Huang, Wen-Chin and Toda, Tomoki},
  booktitle={Proc. ASRU},
  year={2023}
}

@inproceedings{liu2021diffsvc,
  title={{DiffSVC}: A diffusion probabilistic model for singing voice conversion},
  author={Liu, Songxiang and Cao, Yuewen and Su, Dan and Meng, Helen},
  booktitle={Proc. ASRU},
  pages={741--748},
  year={2021},
  organization={IEEE}
}

@inproceedings{liu2021fastsvc,
  title={{FastSVC}: Fast cross-domain singing voice conversion with feature-wise linear modulation},
  author={Liu, Songxiang and Cao, Yuewen and Hu, Na and Su, Dan and Meng, Helen},
  booktitle={Proc. ICME},
  pages={1--6},
  year={2021},
  organization={IEEE}
}

@article{huang2022comparative,
  title={A comparative study of self-supervised speech representation based voice conversion},
  author={Huang, Wen-Chin and Yang, Shu-Wen and Hayashi, Tomoki and Toda, Tomoki},
  journal={IEEE Journal of Selected Topics in Signal Processing},
  volume={16},
  number={6},
  pages={1308--1318},
  year={2022},
  publisher={IEEE}
}

@article{ren2019fastspeech,
  title={Fastspeech: Fast, robust and controllable text to speech},
  author={Ren, Yi and Ruan, Yangjun and Tan, Xu and Qin, Tao and Zhao, Sheng and Zhao, Zhou and Liu, Tie-Yan},
  journal={Advances in neural information processing systems},
  volume={32},
  year={2019}
}

@misc{2020ritsu_enunuv2,
  title = {{[NamineRitsu] Blue (YOASOBI) [ENUNU model Ver.2, Singing DBVer.2 release]}},
  author = {Canon},
  howpublished = {\url{https://www.youtube.com/watch?v=pKeo9IE_L1I}},
  note = {Accessed: 2024.03.14}
}

@misc{natsume,
  title = {{Sota Kirino, Yuuri Natsume Official Website}},
  author = {Kirino, Sota and Natsume, Yuuri},
  howpublished = {\url{https://ksdcm1ng.wixsite.com/njksofficial}},
  note = {Accessed: 2024.03.14}
}

@misc{itako,
  title = {{Tohoku Itako Official Website}},
  author = {Itako},
  howpublished = {\url{https://zunko.jp/itadev/login.php}},
  note = {Accessed: 2024.03.14}
}

@inproceedings{torres2023sslembeddings,
  title={Singer Identity Representation Learning using Self-Supervised Techniques},
  author={Torres, Bernardo and Lattner, Stefan and Richard, Gael},
  booktitle={Proc. ISMIR},
  year={2023}
}

@inproceedings{polyphonic2022midi,
  title={Pseudo-Label Transfer from Frame-Level to Note-Level in a Teacher-Student Framework for Singing Transcription from Polyphonic Music},
  author={Sangeun Kum and Jongpil Lee and Keunhyoung Luke Kim and Taehyoung Kim and Juhan Nam},
  booktitle={Proc. IEEE ICASSP},
  year={2022}
}

@INPROCEEDINGS{2023yongphonememidi,
  author={Yong, Sangeon and Su, Li and Nam, Juhan},
  booktitle={Proc. IEEE ICASSP}, 
  title={A Phoneme-Informed Neural Network Model For Note-Level Singing Transcription}, 
  year={2023},
  volume={},
  number={},
  doi={10.1109/ICASSP49357.2023.10096707}}

@article{hsu2021hubert,
  title={Hubert: Self-supervised speech representation learning by masked prediction of hidden units},
  author={Hsu, Wei-Ning and Bolte, Benjamin and Tsai, Yao-Hung Hubert and Lakhotia, Kushal and Salakhutdinov, Ruslan and Mohamed, Abdelrahman},
  journal={IEEE/ACM Transactions on Audio, Speech, and Language Processing},
  volume={29},
  pages={3451--3460},
  year={2021},
  publisher={IEEE}
}

@inproceedings{
    hsoft-vc-2022,
    author={van Niekerk, Benjamin and Carbonneau, Marc-André and Zaïdi, Julian and Baas, Matthew and Seuté, Hugo and Kamper, Herman},
    booktitle={Proc. IEEE ICASSP}, 
    title={A Comparison of Discrete and Soft Speech Units for Improved Voice Conversion}, 
    year={2022}
}

@INPROCEEDINGS{yoneyama23sifigan,
  author={Yoneyama, Reo and Wu, Yi-Chiao and Toda, Tomoki},
  booktitle={Proc. IEEE ICASSP},
  title={{Source-Filter HiFi-GAN: Fast and Pitch Controllable High-Fidelity Neural Vocoder}},
  year={2023},
  volume={},
  number={},
  doi={10.1109/ICASSP49357.2023.10095298}
}

@article{transformer,
  title={Attention is all you need},
  author={Vaswani, Ashish and Shazeer, Noam and Parmar, Niki and Uszkoreit, Jakob and Jones, Llion and Gomez, Aidan N and Kaiser, {\L}ukasz and Polosukhin, Illia},
  journal={Advances in neural information processing systems},
  volume={30},
  year={2017}
}

@article{ho2020denoising,
  title={Denoising diffusion probabilistic models},
  author={Ho, Jonathan and Jain, Ajay and Abbeel, Pieter},
  journal={Proc. NeurIPS},
  volume={33},
  pages={6840--6851},
  year={2020}
}

@inproceedings{zhou2023vits,
  title={VITS-based Singing Voice Conversion System with DSPGAN post-processing for SVCC2023},
  author={Zhou, Yiquan and Chen, Meng and Lei, Yi and Zhu, Jihua and Zhao, Weifeng},
  booktitle={Proc. IEEE ASRU},
  year={2023}
}

@inproceedings{ning2023vits,
  title={Vits-Based Singing Voice Conversion Leveraging Whisper and Multi-Scale F0 Modeling},
  author={Ning, Ziqian and Jiang, Yuepeng and Wang, Zhichao and Zhang, Bin and Xie, Lei},
  booktitle={Proc. IEEE ASRU},
  year={2023},
}

@INPROCEEDINGS{unsupervised-svc,
  title={{Unsupervised Singing Voice Conversion}},
  author={Nachmani, E. and Wolf, L.},
  booktitle={Proc. Interspeech},
  year={2019}
}

@INPROCEEDINGS{ucd-svc,
  title={{Unsupervised Cross-Domain Singing Voice Conversion}},
  author={Polyak, A. and Wolf, L. and Adi, Y. and Taigman, Y.},
  booktitle={Proc. Interspeech},
  year={2020}
}

@inproceedings{zhou2022hifi,
  title={{HiFi-SVC: Fast High Fidelity Cross-Domain Singing Voice Conversion}},
  author={Zhou, Y. and Lu, X.},
  booktitle={Proc. ICASSP},
  pages={6667--6671},
  year={2022},
}

@inproceedings{
      m4singer,
      title={{M4Singer: A Multi-Style, Multi-Singer and Musical Score Provided Mandarin Singing Corpus}},
      author={L. Zhang and R. Li and S. Wang and L. Deng and J. Liu and Y. Ren and J. He and R. Huang and J. Zhu and X. Chen and Z. Zhao},
      booktitle={Proc. NeruIPS: Datasets and Benchmarks Track},
      year={2022},
    }

@inproceedings{opencpop,
  author={Y. Wang and X. Wang and P. Zhu and J. Wu and H. Li and H. Xue and Y. Zhang and L. Xie and M. Bi},
  title={{OpenCPop: A High-Quality Open Source Chinese Popular Song Corpus for Singing Voice Synthesis}},
  year=2022,
  booktitle={Proc. Interspeech},
  pages={4242--4246},
  doi={10.21437/Interspeech.2022-48}
}

@INPROCEEDINGS{nus48e,
  author={Duan, Z. and Fang, H. and Li, B. and Sim, K. C. and Wang, Y.},
  booktitle={Proc. APSIPA ASC}, 
  title={{The NUS sung and spoken lyrics corpus: A quantitative comparison of singing and speech}}, 
  year={2013},
  volume={},
  number={},
  pages={1-9},
}

@inproceedings{koguchi2020pjs,
  title={{PJS}: Phoneme-balanced Japanese singing-voice corpus},
  author={Koguchi, Junya and Takamichi, Shinnosuke and Morise, Masanori},
  booktitle={Proc. APSIPA ASC},
  pages={487--491},
  year={2020},
  organization={IEEE}
}

@article{ogawa2021tohoku,
  title={Tohoku Kiritan singing database: A singing database for statistical parametric singing synthesis using Japanese pop songs},
  author={Ogawa, Itsuki and Morise, Masanori},
  journal={Acoustical Science and Technology},
  volume={42},
  number={3},
  pages={140--145},
  year={2021},
  publisher={Acoustical Society of Japan}
}

@inproceedings{zen2019libritts,
  author={Heiga Zen and Viet Dang and Rob Clark and Yu Zhang and Ron J. Weiss and Ye Jia and Zhifeng Chen and Yonghui Wu},
  title={{LibriTTS: A Corpus Derived from {LibriSpeech} for Text-to-Speech}},
  year=2019,
  booktitle={Proc. Interspeech},
  pages={1526--1530},
}

@article{sonobe2017jsut,
  title={{JSUT} corpus: free large-scale Japanese speech corpus for end-to-end speech synthesis},
  author={Sonobe, Ryosuke and Takamichi, Shinnosuke and Saruwatari, Hiroshi},
  journal={arXiv preprint arXiv:1711.00354},
  year={2017}
}

@inproceedings{choi2020csd,
  title={Children’s song dataset for singing voice research},
  author={Choi, Soonbeom and Kim, Wonil and Park, Saebyul and Yong, Sangeon and Nam, Juhan},
  booktitle={Proc. ISMIR},
  year={2020}
}

@inproceedings{shen2018taco2,
  title={Natural tts synthesis by conditioning wavenet on mel spectrogram predictions},
  author={Shen, Jonathan and Pang, Ruoming and Weiss, Ron J and Schuster, Mike and Jaitly, Navdeep and Yang, Zongheng and Chen, Zhifeng and Zhang, Yu and Wang, Yuxuan and Skerrv-Ryan, Rj and others},
  booktitle={Proc. IEEE ICASSP},
  pages={4779--4783},
  year={2018}
}

@inproceedings{wu2022_zerosvs,
  author       = {Jui-Te Wu and
                  Jun-You Wang and
                  Jyh-Shing Roger Jang and
                  Li Su},
  title        = {{A unified model for zero-shot singing voice 
                   conversion and synthesis}},
  booktitle    = {{Proceedings of the 23rd International Society for 
                   Music Information Retrieval Conference}},
  year         = 2022,
  pages        = {809-816},
  publisher    = {ISMIR},
  month        = nov,
  venue        = {Bengaluru, India},
  doi          = {10.5281/zenodo.7316786},
  url          = {https://doi.org/10.5281/zenodo.7316786}
}

@article{morise2017harvest,
title = "Harvest: A high-performance fundamental frequency estimator from speech signals",
author = "Masanori Morise",
year = "2017",
doi = "10.21437/Interspeech.2017-68",
pages = "2321--2325",
journal = "Proc. INTERSPEECH",
issn = "2308-457X",
}

@inproceedings{zhou2023bisinger,
  title={BiSinger: Bilingual singing voice synthesis},
  author={Zhou, Huali and Lin, Yueqian and Shi, Yao and Sun, Peng and Li, Ming},
  booktitle={2023 IEEE Automatic Speech Recognition and Understanding Workshop (ASRU)},
  pages={1--8},
  year={2023},
  organization={IEEE}
}

@inproceedings{mcauliffe17_mfa,
  author={McAuliffe, Michael and Socolof, Michaela and Mihuc, Sarah and Wagner, Michael and Sonderegger, Morgan},
  title={{Montreal Forced Aligner: Trainable Text-Speech Alignment Using Kaldi}},
  year=2017,
  booktitle={Proc. Interspeech 2017},
  pages={498--502},
  doi={10.21437/Interspeech.2017-1386}
}

@inproceedings{radford2023whisper,
  title={{Robust speech recognition via large-scale weak supervision}},
  author={Radford, Alec and Kim, Jong Wook and Xu, Tao and Brockman, Greg and McLeavey, Christine and Sutskever, Ilya},
  booktitle={International Conference on Machine Learning},
  pages={28492--28518},
  year={2023},
  organization={PMLR}
}

\end{document}